\documentclass[twocolumn,aps,showpacs,pra,citeautoscript,floatfix,groupedaddress,longbibliography]{revtex4-2}

\usepackage{amssymb}
\usepackage{amsmath}

\usepackage{graphicx,xcolor,hyperref}
\usepackage{mathtools}
\usepackage{natbib}
\usepackage{url}
\usepackage{braket}
\usepackage{relsize}
\usepackage{afterpage}
\newcommand{\argmax}{\mathop{\rm arg~max}\limits}

\usepackage[caption=false,labelformat=simple,position=top]{subfig}

\usepackage[utf8]{inputenc}

\begin{document}

\title{Noisy quantum amplitude estimation without noise estimation}
\author{Tomoki~Tanaka$^{1,2,3}$}
\author{Shumpei~Uno$^{1,4}$}
\author{Tamiya~Onodera$^{1.5}$}
\author{Naoki~Yamamoto$^{1,6}$}
\author{Yohichi~Suzuki$^1$}
\affiliation{$^1$Quantum Computing Center, Keio University, Hiyoshi 3-14-1, Kohoku-ku, Yokohama 223-8522, Japan }
\affiliation{$^2$Mitsubishi UFJ Financial Group, Inc. and MUFG Bank, Ltd.,\\ 2-7-1 Marunouchi, Chiyoda-ku, Tokyo, 100-8388, Japan}
\affiliation{$^3$Graduate School of Science and Technology, Keio University, 3-14-1 Hiyoshi, Kohoku-ku, Yokohama, 223- 8522, Japan}
\affiliation{$^4$Mizuho Research \& Technologies, Ltd, 2-3 Kanda-Nishikicho, Chiyoda-ku, Tokyo, 101-8443, Japan }
\affiliation{$^5$IBM Quantum, IBM Research-Tokyo, 19-21 Nihonbashi Hakozaki-cho, Chuo-ku, Tokyo, 103-8510, Japan}
\affiliation{$^6$Department of Applied Physics and Physico-Informatics, Keio University, Hiyoshi 3-14-1, Kohoku-ku, Yokohama 223-8522, Japan}

\date{\today}

\begin{abstract}
	Many quantum algorithms contain an important subroutine, the quantum
	amplitude estimation.
	As the name implies, this is essentially the parameter estimation problem
	and thus can be handled via the established statistical estimation theory.
	However, this problem has an intrinsic difficulty that the system, i.e.,
	the real quantum computing device, inevitably introduces unknown noise;
	the probability distribution model then has to incorporate many nuisance
	noise parameters, resulting that the construction of an optimal estimator
	becomes inefficient and difficult.
	For this problem, we apply the theory of nuisance parameters (more specifically,
	the parameter orthogonalization method) to precisely compute the maximum
	likelihood estimator for only the target amplitude parameter, by removing the
	other nuisance noise parameters.
	That is, we can estimate the amplitude parameter without estimating the
	noise parameters. 
	We validate the parameter orthogonalization method in a numerical simulation 
	and study the performance of the estimator in the experiment using a real superconducting quantum device. 
\end{abstract}

\maketitle


\section{Introduction}

Quantum computing is expanding its application areas, yet based on a few
fundamental subroutines, e.g., Grover's amplitude amplification operation
\cite{grover1998quantum} and its extension to the quantum amplitude
estimation (QAE) algorithm \cite{brassard2002}.
Actually, we can directly use QAE to do the general Monte Carlo computation
task with quadratically less computational operations compared to any
conventional classical approach \cite{montanaro2015quantum};
moreover, this quantum-enhanced Monte Carlo computation can be applied
to the problem of option pricing and risk calculation in finance
\cite{rebentrost2018quantum,woerner2019quantum,stamatopoulos2020option,martin2021toward,egger2020credit,miyamoto2020reduction,kaneko2021quantum,chakrabarti2021threshold,miyamoto2021bermudan}.

Such a progress of the area is supported by the recent rapid development
of prototypes of real quantum computing devices \cite{huang2020superconducting,stehli2020coherent,haffner2008quantum,he2019two,wang2021minimizing,veldhorst2015two,bruzewicz2019trapped,jurcevic2021demonstration},
some of which provide even a cloud-based worldwide use.
However, they are still in their infancy with several limitations, especially
the noise (decoherence).
Therefore recently we find several elaborated quantum algorithms that could
even run on those noisy quantum devices.
For the case of QAE,
Refs.~\cite{suzuki2020,aaronson2020quantum,grinko2021iterative,nakaji2020fasterae}
provide algorithms that yield an estimator via parallel running of short
Grover operations and postprocessing, yet under the noiseless assumption.
Later, some QAE algorithms that try to improve the estimation performance
by introducing an explicit noise model were presented
\cite{brown2020quantum,tanaka2021amplitude,uno2021modified,giurgica2020low,plekhanov2021variational,giurgica2021low,wang2021minimizing}.
In particular, the depolarizing noise is often assumed \cite{tanaka2021amplitude,uno2021modified,giurgica2020low,giurgica2021low},
meaning that we study the model probability distribution
$P(\mathbf{h} ; \theta, \beta)$ where $\theta$ is the amplitude parameter
and $\beta$ is the noise parameter.

The difficulty of this approach lies in the fact that it is impossible to
have a complete parametric model.
The depolarizing noise model may explain many of imperfection observed in
experiments, but there always exist remaining and unidentifiable noise.
Hence we need a more complicated parametric model
$P(\mathbf{h} ; \theta, \boldsymbol{\beta})$ with noise parameters
$\boldsymbol{\beta}=(\beta_1, \beta_2, \ldots, \beta_M)$, to better
characterize such unknown noise sources.
However, naively this approach forces us to construct an estimator for
all those parameters;
for instance, the maximum likelihood (ML) estimator requires us to solve
the corresponding $(M+1)$-dimensional optimization problem, which is in
general hard to solve especially when $M$ is large.
Note that this is a general issue in all statistical estimation problems
for realistic systems.
The theory of nuisance parameters (e.g., \cite{cox1987parameter}) offers
a method for dealing with this issue;
that is, we can apply the {\it parameter orthogonalization method} to
efficiently and precisely construct an estimator of only $\theta$, without
respect to the nuisance parameters $\boldsymbol{\beta}$.
That is, we can estimate the amplitude parameter without estimating the
noise parameters.
Note that in the infinite limit $M \to \infty$, we will have a 
{\it semi parametric model}, with $\boldsymbol{\beta}$ replaced by a function 
$\beta_t$ which does not have a specific form and thus can capture infinitely 
many noise effects implicitly. 
The semi parametric estimation theory (e.g., \cite{begun1983information})
provides a method for possibly removing even such an infinite dimensional
function for efficiently constructing an estimator of $\theta$.
Notably, recently we find quantum versions of the parameter orthogonalization 
method \cite{suzuki2020nuisance,suzuki2020quantum} and the semi parametric 
theory \cite{tsang2020quantum,cimini2021semiparametric}, although this paper
focuses on the use of classical theory.

In this paper we first apply the theory of nuisance parameters to efficiently
compute the amplitude parameter in the QAE problem, where the model can cover
a wide range of noise sources, including the depolarization noise.
Below is the summary of the results.
	{\bf (i)}
Based on the experimental results obtained using a real quantum device, we
formulate a valid parametric model $P(\mathbf{h} ; \theta, \boldsymbol{\beta})$,
which is thought to degenerate (i.e., the corresponding Fisher information matrix
is degenerate) and the theory of nuisance parameters cannot be applied.
To circumvent this degeneracy, our idea is to introduce an ancillary
quantum circuit that produces a similar but strictly different distribution
to $P$ such that the combined joint probability distribution is no longer
degenerate.
	{\bf (ii)}
We apply the parameter orthogonalization method to the above-mentioned
joint probability distribution, which requires us to solve a multi-variable
differential equation; importantly, we derive an analytic solution of
this equation.
As a result, the ML estimator $\hat{\theta}_{\rm ML}$ can be obtained
almost exactly by solving a one-dimensional optimization problem;
in other words, we can have the estimator without spending any effort
to estimate the noise parameters $\boldsymbol{\beta}$.
	{\bf (iii)}
In a toy example we numerically validate the parameter orthogonalization
method in the non-asymptotic regime, meaning that the corresponding
likelihood equation is actually almost independent to the nuisance
parameters even with relatively small number of measurement.
Then we show that, in the experiment on a real superconducting quantum
device, the ML estimator works pretty well even under un-identifiable
realistic noise, implying the effectiveness of our strategy for the QAE
problem.

This paper is organized as follows.
Section~\ref{sec:Preliminaries} gives a summary of the existing QAE methods and the theory of nuisance
parameters.
Then in Sec.~\ref{sec:MLAE using parameter orthogonalization method} we describe the above-mentioned results (i) and (ii).
Section~\ref{sec:Numerical and experimental demonstrations} gives an experimental demonstration to show the effectiveness of the proposed
method, i.e., the result (iii) is presented.
Finally Sec.~\ref{sec:Conclusion} concludes the paper.


\section{Preliminaries}
\label{sec:Preliminaries}

\subsection{Quantum amplitude estimation via maximum likelihood method}

Here we describe the QAE problem and the ML method (MLAE) in the ideal noiseless case.
First, the parameter of interest, $\theta \in [0,\pi/2]$, is encoded into the
amplitude of a quantum state, via the operator $\mathcal{A}$ as follows:
\begin{equation}
	\ket{\Psi}_{n+1}
	= \mathcal{A}\ket{0}_{n+1} =
	\sin\theta \ket{\tilde{\Psi}_1}_n\ket{1}
	+\cos\theta \ket{\tilde{\Psi}_0}_n\ket{0},
	\label{eq:defA}
\end{equation}
where $\ket{\tilde{\Psi}_1}_n$ and $\ket{\tilde{\Psi}_0}_n$ are fixed $n$-qubit
quantum states, and
$\ket{0}_{n}$ denotes $\ket{0}^{\otimes n}$.
The amplitude amplification operator (or Grover operator) $\mathcal{G}$ is
defined as
\begin{equation*}
	\mathcal{G} = - \mathcal{A} \mathcal{U}_0 \mathcal{A}^\dagger \mathcal{U}_f,
\end{equation*}
where $U_0$ and $U_f$ are defined as
\begin{equation*}
	\begin{split}
		\mathcal{U}_0 &= I_{n+1}-2\ket{0}_{n+1}\bra{0}_{n+1}, \\
		\mathcal{U}_f &= I_{n+1}-2I_{n}\otimes\ket{1}\bra{1}
		= I_{n}\otimes \sigma_z.
	\end{split}
\end{equation*}
Here $I_{n}$ is the $n$-dimensional identity matrix.
To estimate $\theta$, we take the parallel strategy \cite{suzuki2020};
that is, we apply $\mathcal{G}$ on the initial state $m_k$ times with
several non-negative integers $m_k$ $(k=1,\ldots,M)$ and then combine
the measurement result performed on $\mathcal{G}^{m_k} \ket{\Psi}_{n+1}$
for all $k$ to construct an ML estimator.
For instance, if $m_k=k$, the estimator achieves the error $\epsilon$
using ${\cal O}(\epsilon^{-4/3})$ queries in the noiseless case, while
${\cal O}(\epsilon^{-2})$ is necessary via any classical means.
More specifically, we have
\begin{align}
	\begin{split}
		\mathcal{G}^{m_k} \ket{\Psi}_{n+1}
		=& \sin((2m_k+1)\theta)\ket{\tilde{\Psi}_1}_n\ket{1}\\
		&+ \cos((2 m_k+1)\theta)\ket{\tilde{\Psi}_0}_n\ket{0}.
		\label{eq:Qm}
	\end{split}
\end{align}
For this quantum state, we measure the last qubit with the measurement basis $\{\ket{0}, \ket{1}\}$;
then the probability to obtain the result ``1'' is given by
\begin{equation}
	\label{ideal hitting prob 1}
	p_{\theta}^{(k)} = {\mathbb P}(\{h_k=1\}) = \sin^2((2m_k+1)\theta),
\end{equation}
where $h_k\in\{0,1\}$ is a binary random variable.
Denoting the collection of random variables for all $m_k$ as
$\mathbf{h} = (h_1,\cdots, h_{M})\in\{0,1\}^M$, the joint probability to
have $\mathbf{h}$ is
\begin{equation}
	P(\mathbf{h} ; \theta) = \prod_{k={1}}^M (p_{\theta}^{(k)})^{h_k} (1-p_{\theta}^{(k)})^{1-h_k}.
\end{equation}
Now we make $N_{\rm shot}$ measurements (shots) for a fixed $m_k$ and
collect all the result to
$\mathbf{H}=\{\mathbf{h}^1,\cdots,\mathbf{h}^{N_{\rm{shot}}} \}$;
that is, $\mathbf{H}$ is the set of samples from the distribution
$P(\mathbf{h} ; \theta)$ or equivalently the set of realizations of random
variable $\mathbf{h}$.
Then the ML estimator for $\theta$ is obtained as
\begin{equation}
	\hat{\theta}_{\rm ML}
	=\argmax_{\theta} L(\mathbf{H};\theta)
	=\argmax_{\theta}\ln L(\mathbf{H};\theta),
	\label{maximize_L}
\end{equation}
where $L(\mathbf{H}; \theta)$ is the likelihood function
\begin{equation}
	L(\mathbf{H}; \theta)
	= \prod_{j={1}}^{N_{\rm{shot}}} P(\mathbf{h}^j ; \theta).
	\label{likelihood}
\end{equation}

Now the estimation error of $\theta$ via any estimator $\hat{\theta}$ 
can be evaluated by the Cram\'{e}r--Rao inequality
\begin{equation}
	\label{CR inequality}
	\epsilon^2 = \mathbb{E}\big[ (\theta - \hat{\theta} )^2 \big]
	\ge \frac{1}{N_{\textrm{shot}}}J(\theta)^{-1},
\end{equation}
where $J(\theta)$ is the Fisher information, defined as
\begin{equation}
	J(\theta)
	=\mathbb{E}\Big[
		\Big( \frac{\partial}{\partial \theta}\ln P(\mathbf{h};\theta) \Big)^2
		\Big].
	\label{Fisher info def}
\end{equation}
Here $\mathbb{E}[\cdots]$ represents the expectation over random 
variables $\mathbf{h}$. 
The strength of the ML estimate $\hat{\theta}_{\rm ML}$ is that it 
can asymptotically achieve the lower bound of Cram\'{e}r--Rao 
inequality.

We now discuss the case where the system is under a noisy environment.
As mentioned in Sec.~I, one may introduce a typical noise model and
consider the corresponding parametric probability distribution for executing
the above estimation procedure.
For instance, in Ref.\cite{tanaka2021amplitude}, the depolarizing noise with
noise parameter $\beta$ is assumed, and the resulting classical probability
distribution $P(\mathbf{h};\theta,\beta)$ is used to construct the ML
estimator for both $(\theta,\beta)$ by maximizing the two-dimensional
likelihood function
$L(\mathbf{H}; \theta, \beta)
	= \prod_{j={1}}^{N_{\rm{shot}}} P(\mathbf{h}^j ; \theta ,\beta)$.
However, in practice there are many noise sources other than the
depolarizing noise, such as the dephasing noise, meaning that there
must exist a gap between $P(\mathbf{h};\theta,\beta)$ and the unknown
true distribution.
To decrease this gap, we could consider a more complicated parametric
probability distribution composed of several possible noise sources,
$P(\mathbf{h};\theta,\boldsymbol{\beta})$ characterized by the vector
of noise parameters $\boldsymbol{\beta}$.
Although the expressibility becomes higher and consequently, the gap
would become small, this approach must force us to maximize the
multi-dimensional (non-convex) function
$L(\mathbf{H};\theta, \boldsymbol{\beta})$ with respect to
$(\theta, \boldsymbol{\beta})$ to have the ML estimator $\hat\theta_{\rm ML}$.
Then, if particularly the size of $\boldsymbol{\beta}$ is large, we could fail
to exactly maximize $L(\mathbf{H};\theta, \boldsymbol{\beta})$ and
consequently only have a suboptimal estimator that can largely differ
from the ML estimator, which as a result degrades the estimation performance
on $\theta$.
The theory of nuisance parameters described in the next section can
be used to resolve this issue.

\subsection{Theory of nuisance parameters}
\label{Theory of nuisance parameters}

As mentioned above, the main difficulty in the multi-parameter estimation
lies in the hardness to solve the multi-dimensional optimization problem.
Fortunately, the theory of nuisance parameters provides a condition such
that this optimization problem boils down to a one-dimensional one with
respect to only $\theta$.
This method is called the parameter orthogonalization method, which allows
us to separate the parameter of interest and the nuisance parameters.

Here we describe the essential idea of the parameter orthogonalization method;
see \cite{cox1987parameter} for a more detailed description.
Let us consider a general probability distribution
$p(x;\theta, \boldsymbol{\beta})$ where $\theta$ is the parameter of interest
and $\boldsymbol{\beta}=(\beta_1,\ldots,\beta_M)$ are nuisance parameters.
Now let us consider the following transformation of parameters:
\begin{equation*}
	\theta = \theta(\xi_1)=\xi_1, ~
	\beta_k = \beta_k(\xi_1,\xi_2,\cdots,\xi_{M+1}).
\end{equation*}
The new parameters $\boldsymbol{\xi}=(\xi_1,\xi_2, \ldots, \xi_{M+1})$ are to
be determined so that the new Fisher information matrix $J_{\xi}$ satisfies
$(J_{\xi})_{1,k}=0$ (that is, the $(1, k)$ element of $J_{\xi}$ is zero)
for all  $k=2,3,\ldots,M+1$. As shown in Appendix, this requirement indeed holds if 
$\beta_k(\boldsymbol{\xi})$ is the solution
of the differential equation
\begin{eqnarray}
	\label{PDF_org}
	J_{1,i}+\sum_{k=2}^{M+1}J_{i,k}
	\frac{\partial \beta_{k-1}}{\partial \xi_1}=0,
\end{eqnarray}
for all $i=2,3, \ldots, M+1$.
Note that in general Eq.~\eqref{PDF_org} does not have a unique solution,
as will be demonstrated in the next section.
The benefit of the parameter orthogonalization condition $(J_{\xi})_{1,k}=0,
	\forall k\geq 2$ is clear; that is, under certain regular conditions, it can
be rewritten as
\begin{align*}
	0=(J_{\xi})_{1,k}
	 & ={\mathbb E}\Big[ \frac{\partial \ln p(x;\boldsymbol{\xi})}
		{\partial \xi_1}
		\frac{\partial \ln p(x;\boldsymbol{\xi})}
		{\partial \xi_k}
		\Big]                                                                  \\
	 & =-{\mathbb E}\Big[ \frac{\partial^2}{\partial \xi_1 \partial \xi_k}
		\ln p(x;\boldsymbol{\xi}) \Big]                                        \\
	 & \approx -\frac{1}{N_x}\sum_g
	\frac{\partial^2}{\partial \xi_1 \partial \xi_k}
	\ln p(x_g;\boldsymbol{\xi})                                            \\
	 & = -\frac{1}{N_x} \frac{\partial^2}{\partial \xi_1 \partial \xi_k}
	\ln L(\boldsymbol{x};\boldsymbol{\xi}),
\end{align*}
meaning that the likelihood equation with respect to $\theta=\xi_1$, i.e.,
$(\partial/\partial \xi_1)\ln L(\boldsymbol{x};\boldsymbol{\xi}) = 0$, does
not depend on $\xi_k,~\forall k\geq 2$.
This means that, in the asymptotic limit where the number of samples $N_x$
becomes infinite, the critical points of $\theta$ do not depend on
$\xi_k,~\forall k\geq 2$
and thus the ML estimator for $\theta$ can be
computed by simply solving the one-dimensional optimization problem
$\max_{\xi_1} \ln L(\boldsymbol{x};\xi_1, \bar{\xi}_2, \ldots, \bar{\xi}_{M+1} )$ 
with roughly chosen $\bar{\xi}_k~\forall k\geq 2$.
In Sec.~IV A, we demonstrate that this parameter orthogonalization is almost satisfied in our QAE problem, even when the number of samples (measurements in our case) is finite.
Finally note that the Cram\'{e}r-Rao lower bound (CRLB) on $\theta$, i.e.,
$(J_{\xi}^{-1})_{1,1}=(J_{\xi})_{1,1}^{-1}$, is the same as $(J^{-1})_{1,1}$, which is achieved
by the multi-parameter ML estimator for $(\theta, \boldsymbol{\beta})$.
Also we remark that, when there are multiple parameters of interest, 
there is no general strategy for parameter orthogonalization.


\section{MLAE using the parameter orthogonalization method}
\label{sec:MLAE using parameter orthogonalization method}

\subsection{Ancillary Grover operator}

\begin{figure}[tb]
	\begin{center}
		\includegraphics[width=\linewidth]{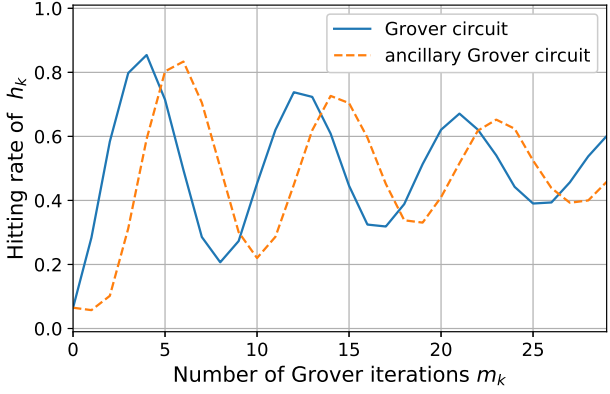}
		\caption{Experimental result of the probability of hitting ``1", as a
			function of the number of Grover iterations, $m_k=k$.
			The (blue) solid line is obtained from the Grover circuit of \eqref{eq:Qm},
			and the (orange) dashed line is from the ancillary Grover circuit of \eqref{eq:ancillary_circuit}.
			This experiment used the first and the fourth qubits of 
			``ibm\_kawasaki," and $N_{\rm{shot}}=8192$ for both 
			circuits. 
		}
		\label{fig:exp_oscillation}
	\end{center}
\end{figure}

We begin with defining a parametric distribution model under the unknown
environment noise.
Our basis is on the experimental result of the probability of hitting 1
as a function of the number of Grover iterations $m_k$, corresponding to
Eq.~\eqref{ideal hitting prob 1} in the ideal case.
The experiment was conducted using the IBM quantum device ``ibm\_kawasaki'',
and the result is shown in the (blue) solid line in 
Fig.~\ref{fig:exp_oscillation}.
This would suggest that we take a decayed oscillation
\begin{equation}
	\label{depolarization case}
	p_{\theta, \kappa}^{(k)}
	= \frac{1}{2}
	- \frac{1}{2} \mathrm{e}^{- \kappa m_k} \cos(2(2m_k+1)\theta).
\end{equation}
Indeed this can be derived by assuming the depolarizing noise with strength
$\kappa$; see \cite{tanaka2021amplitude}.
Note that if $\kappa=0$, this is exactly Eq.~\eqref{ideal hitting prob 1}.
However, in reality there must exist some noise sources other than the
depolarizing noise.
Hence we introduce the following generalized model:
\begin{align}
	p_{\theta, \beta_k}^{(k)}
	= \frac{1}{2} - \frac{1}{2} \beta_k \cos(2(2m_k+1)\theta).
	\label{eq:proposedprobability}
\end{align}
That is, a different type of noise, which is not necessarily the depolarizing
noise, can be added to the system, depending on the number of iteration $m_k$.
Note that Eq.~\eqref{eq:proposedprobability} can be originated to the
continuous-time model
$p(t; \theta, \beta(t)) = (1-\beta(t) \cos(2(2t+1)\theta))/2$
with $t$ the running time of the Grover operator;
this is a semi parametric model with unknown function $\beta(t)$, which in
our case is reduced to a finite-dimensional parametric model as only the
finite number of sampling is performed.

The joint probability taken in the parallel strategy is given by
\begin{equation}
	P(\mathbf{h} ; \theta,\boldsymbol{\beta}) = \prod_{k={1}}^M
	[p_{\theta, \beta_k}^{(k)}]^{h_k}
	[1-p_{\theta, \beta_k}^{(k)}]^{1-h_k},
\end{equation}
and the likelihood function becomes
\begin{equation}
	L(\mathbf{H}; \theta, \boldsymbol{\beta})
	= \prod_{j=1}^{N_{\rm{shot}}}P(\mathbf{h}^j ; \theta,\boldsymbol{\beta}),
	\label{noisy likelihood}
\end{equation}
where the meaning of $h_k$, $\mathbf{h}^j$, and $\mathbf{H}$ are the
same as before.
Therefore, the model distribution is parametrized by $\theta$ and the
nuisance parameters $\boldsymbol{\beta}=(\beta_1,\ldots, \beta_M)$.

As mentioned before, we may fail to perfectly solve the optimization problem
$\max_{\theta, \boldsymbol{\beta}} L(\mathbf{H};\theta, \boldsymbol{\beta})$
and obtain the ML estimator $\hat{\theta}_{\rm ML}$, especially when $M$ is
large;
this is indeed the motivation to apply the parameter orthogonalization method
to remove $\boldsymbol{\beta}$.
However, we cannot do this in a straightforward way, because, in this case,
the model distribution is not regular; in fact, the number of outputs is
$M$ while that of parameters is $M+1$, leading to the Fisher information
matrix being degenerate.

Therefore we need to introduce an ancillary system that yields additional
outputs, while keeping the number of noise parameters to recover the
regularity of the model.
In other words, we need a quantum circuit that is governed by the same
noise as that on the entire Grover operation $\mathcal{G}^m$ and yet
produces a different measurement outcome.
For this purpose, we introduce the following ``unamplified” operator
$\mathcal{R}$:
\begin{equation}
	\mathcal{R} := - \mathcal{A} U_0 \mathcal{A}^\dagger I_{n+1}.
\end{equation}
This operator is obtained by replacing $\mathcal{U}_f=I_{n}\otimes\sigma_z$
by the identity operator $I_{n+1}$ in the Grover operator
$\mathcal{G}= - \mathcal{A} \mathcal{U}_0 \mathcal{A}^\dagger \mathcal{U}_f$.
Note that, in quantum devices operated with the computational-basis
measurement (i.e., measurement in the $Z$ basis), such as the current
IBM superconducting devices, the $Z$ gate is implemented via the frame change;
that is, no pulselike operation is performed on the system
\cite{mckay2017efficient}.
This technique has been applied to other types of quantum devices, such as
NMR~\cite{knill2000algorithmic} and trapped ions~\cite{knill2008randomized}.
Hence, it might be acceptable to assume that the operators $\mathcal{G}$
and $\mathcal{R}$ are subjected to the same noise.
This basic assumption is supported in an experiment, as will be explained
below.

Based on the above-introduced $\mathcal{R}$, we define the ancillary
Grover circuit as follows;
that is, we replace the last operation $\mathcal{G}$ of the entire Grover
operations by $\mathcal{R}$.
Then the final state in the ideal noiseless case is given by
\begin{equation}
	\begin{split}
		\mathcal{R} \mathcal{G}^{m_k-1} \ket{\Psi}_{n+1}
		=& ( - \mathcal{A} U_0 \mathcal{A}^\dagger) (
		- \mathcal{A} \mathcal{U}_0 \mathcal{A}^\dagger \mathcal{U}_f)\mathcal{G}^{m_k-2} \ket{\Psi}_{n+1}\\
		=& \mathcal{U}_f \mathcal{G}^{m_k-2} \ket{\Psi}_{n+1}.
	\end{split}
	\label{eq:ancillary_circuit}
\end{equation}
For the realistic noisy case, the final state will be affected by the
same noise through the above $m_k$ iterations, as that of $\mathcal{G}^{m_k}\ket{\Psi}_{n+1}$.
Hence, the probability of getting 1 by measuring the last qubit of
the circuit $\mathcal{R}\mathcal{G}^{m_k-1}$ under noise is, according to
Eq.~\eqref{eq:proposedprobability} and the fact that it is not affected
by the operator $\mathcal{U}_f$, given by
\begin{equation}
	\begin{split}
		q_{\theta,\beta_k}^{(k)}
		&= \frac{1}{2} - \frac{1}{2} \beta_k \cos(2(2(m_k-2)+1)\theta) \\
		&= \frac{1}{2} - \frac{1}{2} \beta_k \cos(2(2m_k -3)\theta).
		\label{eq:probabilityR}
	\end{split}
\end{equation}
This is the phase-delayed oscillation of Eq.~\eqref{eq:proposedprobability},
and this delay can be clearly seen in Fig.~\ref{fig:exp_oscillation}, 
showing with the (orange) dashed line the actual hitting ratio of 1 in the 
experiment.
This result supports our assumption that the two circuits $\mathcal{G}^{m_k}$
and $\mathcal{R}\mathcal{G}^{m_k-1}$ are affected by the same noise.

The joint probability for ancillary circuits is given by
\begin{equation}
	Q(\boldsymbol{\ell} ; \theta,\boldsymbol{\beta}) = \prod_{k={1}}^M
	[q_{\theta, \beta_k}^{(k)}]^{\ell_k}
	[1-q_{\theta, \beta_k}^{(k)}]^{1-\ell_k},
\end{equation}
and the likelihood function is constructed as
\begin{equation}
	\begin{split}
		L(\mathbf{H},\mathbf{L} ; \theta, \boldsymbol{\beta})
		= \prod_{i=1}^{N_{\rm{shot}}}P(\mathbf{h}^i ; \theta,\boldsymbol{\beta})
		\prod_{j=1}^{N_{\rm{shot}}'}Q(\mathbf{\boldsymbol{\ell}}^j ; \theta,\boldsymbol{\beta}),
	\end{split}
	\label{eq:proposedlikelihood}
\end{equation}
where
$\mathbf{L}=\{\boldsymbol{\ell}^1,\cdots, \boldsymbol{\ell}^{N_{\rm{shots}}'}\}$
is the set of measurement results sampled from the binary random variables $\boldsymbol{\ell}=(\ell_1,\cdots,\ell_M)$, which follow the probability
distributions of ancillary Grover circuits.
Recall that the ancillary Grover circuit $\mathcal{R}\mathcal{G}^{m_k-1}$
is executed for $N_{\rm{shots}}'$ times.
Also with the help of an ancillary circuit, the corresponding Fisher information
matrix can be invertible, meaning that certainly we now have the regular
model to which the parameter orthogonalization method is applicable.


\subsection{Orthogonalized parameters in MLAE}
\label{sec:Orthogonalized parameters in MLAE}

We can now apply the parameter orthogonalization method described in
Sec.~\ref{Theory of nuisance parameters} to our problem.
The goal is to find the transformation such that the differential equation \eqref{PDF_org} is satisfied.
For simplicity, we set $N_{\rm shot}=N_{\rm shot}'$.
In this setting, Eq.~\eqref{PDF_org} is reduced to
\begin{widetext}
	\begin{eqnarray*}
		\frac{\partial \beta_k(\boldsymbol{\xi})}{\partial \xi_1}\left(\frac{A_k^p(\xi_1)}{1-A_k^p(\xi_1)\beta_{k}^2(\boldsymbol{\xi})}+\frac{A_k^q(\xi_1)}{1-A_k^q(\xi_1)\beta_{k}^2(\boldsymbol{\xi})}\right)
		=-\frac{\beta_k(\boldsymbol{\xi})}{2}\left(\frac{1}{1-A_k^p(\xi_1)\beta_{k}^2(\boldsymbol{\xi})}\frac{\partial A_k^p}{\partial \xi_1}+\frac{1}{1-A_k^q(\xi_1)\beta_{k}^2(\boldsymbol{\xi})}\frac{\partial A_k^q}{\partial \xi_1}\right),
	\end{eqnarray*}
	for all $k=1, 2, \cdots, M$.
	Here we defined 
	\begin{eqnarray*}
		A_k^p(\xi_1)=\cos^2(2 (2m_k+1) \xi_1), ~~
		A_k^q(\xi_1)&=&\cos^2(2 (2m_k-3)\xi_1).
	\end{eqnarray*}
	Then we can prove that, for all $k=1, 2, \cdots, M$, the function  $\beta_k(\boldsymbol{\xi})$
	satisfies
	\begin{equation}
		\{1-A_k^p(\xi_1)\beta_k^2(\boldsymbol{\xi})\}\{1-A_k^q(\xi_1)\beta_k^2(\boldsymbol{\xi})\}= c_k(\boldsymbol{\xi}'),
		\label{quartic_eq}
	\end{equation}
	where $c_k$ are arbitrary functions of the nuisance parameters
	$\boldsymbol{\xi}'=(\xi_2,\cdots,\xi_{M+1})$.
	Equation \eqref{quartic_eq} can be solved as
	\begin{eqnarray}
		\beta_k^2(\boldsymbol{\xi})
		=\frac{A_k^p(\xi_1)+A_k^q(\xi_1)\pm\sqrt{(A_k^p(\xi_1)+A_k^q(\xi_1))^2-4A_k^p(\xi_1)A_k^q(\xi_1)(1-c_k(\boldsymbol{\xi}'))}}{2 A_k^p(\xi_1) A_k^q(\xi_1)}. \label{sol_quartic_eq}
	\end{eqnarray}
\end{widetext}
If $\beta_k(\boldsymbol{\xi})$ is a real function of
$\boldsymbol{\xi}$ and $\beta_k(\boldsymbol{\xi})\geq 0$,
the following condition needs to be satisfied;
\begin{eqnarray*}
	1-\frac{(A_k^p(\xi_1)+A_k^q(\xi_1))^2}{4A_k^p(\xi_1)A_k^q(\xi_1)}
	\leq c_k(\boldsymbol{\xi}') \leq 1.
\end{eqnarray*}
Thus, since $(A_k^p(\xi_1)+A_k^q(\xi_1))^2/(4 A_k^p(\xi_1) A_k^q(\xi_1))\geq 1$,
all solutions of $\beta_k(\boldsymbol{\xi})$ are real if
$0 \leq c_k(\boldsymbol{\xi}') \leq 1$.
Moreover, because {$0\leq p^{(k)}_{\theta,\beta_k}\leq 1$} and
$0\leq q^{(k)}_{\theta,\beta_k}\leq 1$, the following conditions need to be
satisfied for all $k$:
\begin{eqnarray*}
	0\leq A_k^p(\xi_1)\beta_k^2(\boldsymbol{\xi})\leq 1, ~~
	0\leq A_k^q(\xi_1)\beta_k^2(\boldsymbol{\xi})\leq 1.
\end{eqnarray*}
The relevant solution of $\beta_k(\boldsymbol{\xi})$ is thus
\begin{widetext}
	\begin{eqnarray}
		\beta_k(\boldsymbol{\xi})
		=\sqrt{\frac{A_k^p(\xi_1)+A_k^q(\xi_1)-\sqrt{(A_k^p(\xi_1)+A_k^q(\xi_1))^2-4A_k^p(\xi_1)A_k^q(\xi_1)(1-c_k(\boldsymbol{\xi}'))}}{2 A_k^p(\xi_1) A_k^q(\xi_1)}}.
		\label{sol_quartic_eq_final}
	\end{eqnarray}
\end{widetext}
Recall from the theory of nuisance parameters shown in Sec.~II~B that,
although $c_k(\boldsymbol{\xi}')$ appearing in Eq.~\eqref{sol_quartic_eq_final}
is an arbitrary function satisfying $0\leq c_k(\boldsymbol{\xi}')\leq 1$,
it does not affect the estimation of $\theta=\xi_1$ in the asymptotic limit of
a large number of measurements.
Therefore by substituting $\beta_k(\boldsymbol{\xi})$ with roughly
chosen $c_k(\boldsymbol{\xi}')$ into the likelihood function
\eqref{eq:proposedlikelihood} and solving the one-dimensional maximization
problem with respect to $\theta=\xi_1$, we can efficiently and almost
correctly compute the ML estimator for $\theta$. 
Lastly note that, if we eventually need to use a numerical solver for 
the differential equation \eqref{PDF_org}, this means that the proposed 
method requires an additional computational resources, which has to be 
carefully compared to that of the multi-dimensional optimizer for the 
likelihood function \eqref{eq:proposedlikelihood}. 
Moreover, such a numerical procedure may easily cause an error to the 
solutions $\beta_k(\boldsymbol{\xi})$ and the resulting likelihood function 
to be maximized with respect to $\theta$. 
Therefore, the fact that we have successfully obtained the analytic solution 
\eqref{sol_quartic_eq_final} is essentially important for the parameter 
orthogonalization method to gain the genuine computational advantage over 
the naive multi-parameter ML method.


\section{Numerical and experimental demonstrations}
\label{sec:Numerical and experimental demonstrations}

In this section we study the performance of the proposed method in both
numerical simulation and an experiment on a real superconducting quantum
device.

\subsection{Numerical demonstration}

First we numerically validate the parameter orthogonalization method in the
non-asymptotic regime where the number of samples (measurement), i.e.,
$N_{\rm shot}$, is finite.
More specifically, we study the solution of our likelihood equation with
respect to $\theta=\xi_1$, i.e.,
\begin{equation}
	\label{likelihood eq Section IV}
	\frac{\partial}{\partial \theta}
	\ln L(\mathbf{H},\mathbf{L} ; \theta, \boldsymbol{\beta}) = 0,
\end{equation}
with $\beta_k(\boldsymbol{\xi})$ given by Eq.~\eqref{sol_quartic_eq_final};
then we will see if those solutions almost do not depend on
$\xi_k$ or equivalently $c_k(\boldsymbol{\xi}')$ for all $k\geq 2$,
even when $N_{\rm shot}$ is relatively a small number.
In fact, in the quantum computation scenario, $N_{\rm shot}$ should be kept
as low as possible, hence this analysis is important.

The parameters of this numerical experiment are chosen as follows:
The target value is $\theta_* = 0.35$.
The number of measurements (shots) is $N_{\rm{shot}}=50$ for both the Grover
and the ancillary Grover circuits.
The amplification schedule is $m_k = 2^{k-1}~(k=1,2,\ldots,8)$.
The true probability distribution is Eq.~\eqref{depolarization case} with
$\kappa=0.01$, which is used to generate the data; that is, only the
depolarizing noise is added to the system, meaning that our parametric model
can represent this true distribution by properly choosing $\beta_k$.
Finally, as for the free parameters $\{c_1,\ldots,c_8\}$ given in
Eq.~\eqref{sol_quartic_eq}, we consider the following two cases:
\begin{eqnarray*}
	{\mathit 1:} & \{0.844, 0.134, 0.956, 0.238, 0.236, 0.623, 0.793, 0.324 \}, \\
	{\mathit 2:} & \{0.571, 0.452, 0.475, 0.259, 0.107, 0.965, 0.362, 0.522 \}.
\end{eqnarray*}

\begin{figure}
	\begin{minipage}{\linewidth}
		\begin{center}
			\subfloat[Case ${\mathit 1}$\label{fig:loglikelihood0}]{ \includegraphics[width=\linewidth]{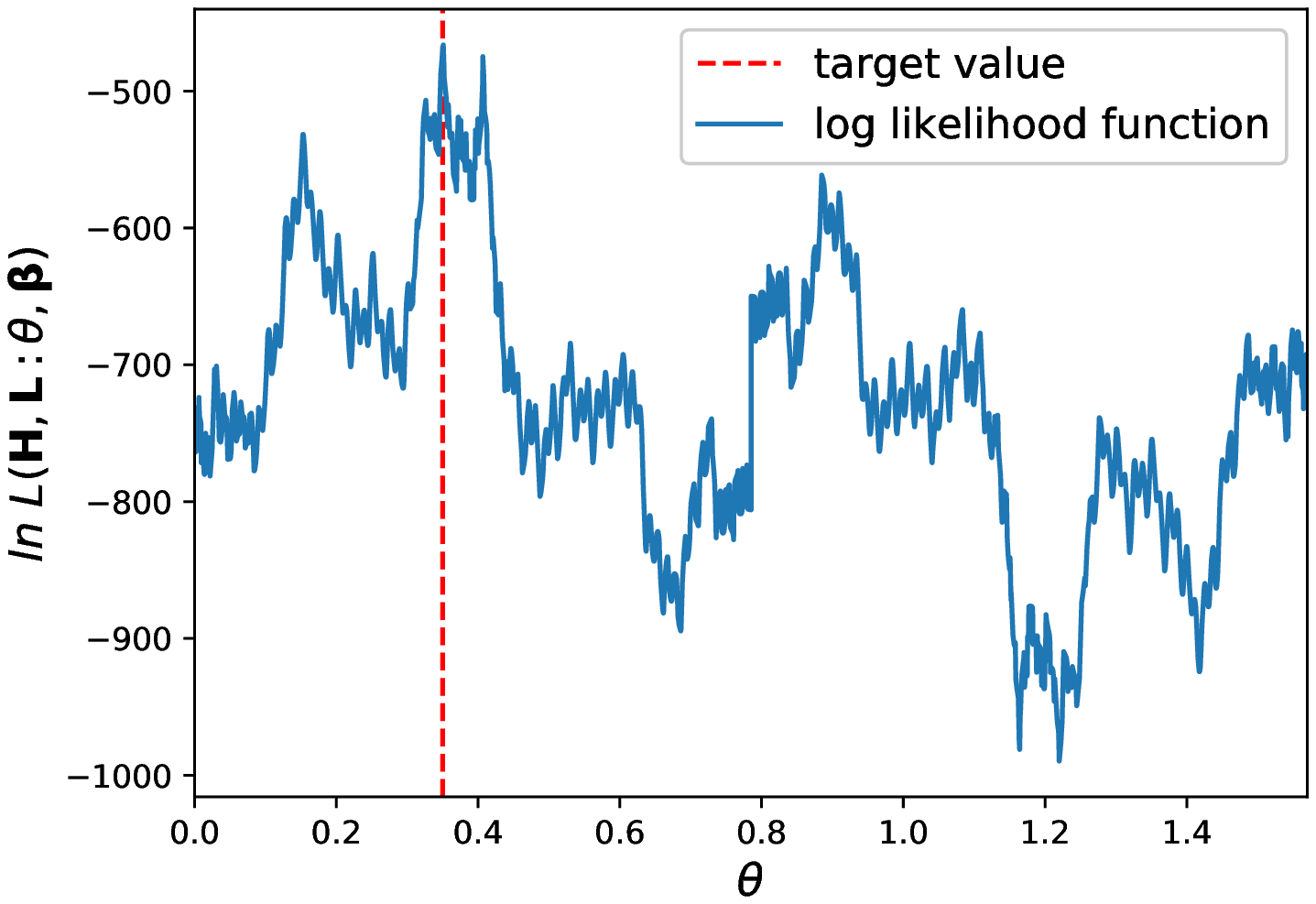} }
		\end{center}
	\end{minipage}\vspace{0.5cm}\\
	\begin{minipage}{\linewidth}
		\begin{center}
			\subfloat[Case ${\mathit 2}$\label{fig:loglikelihood1}]{ \includegraphics[width=\linewidth]{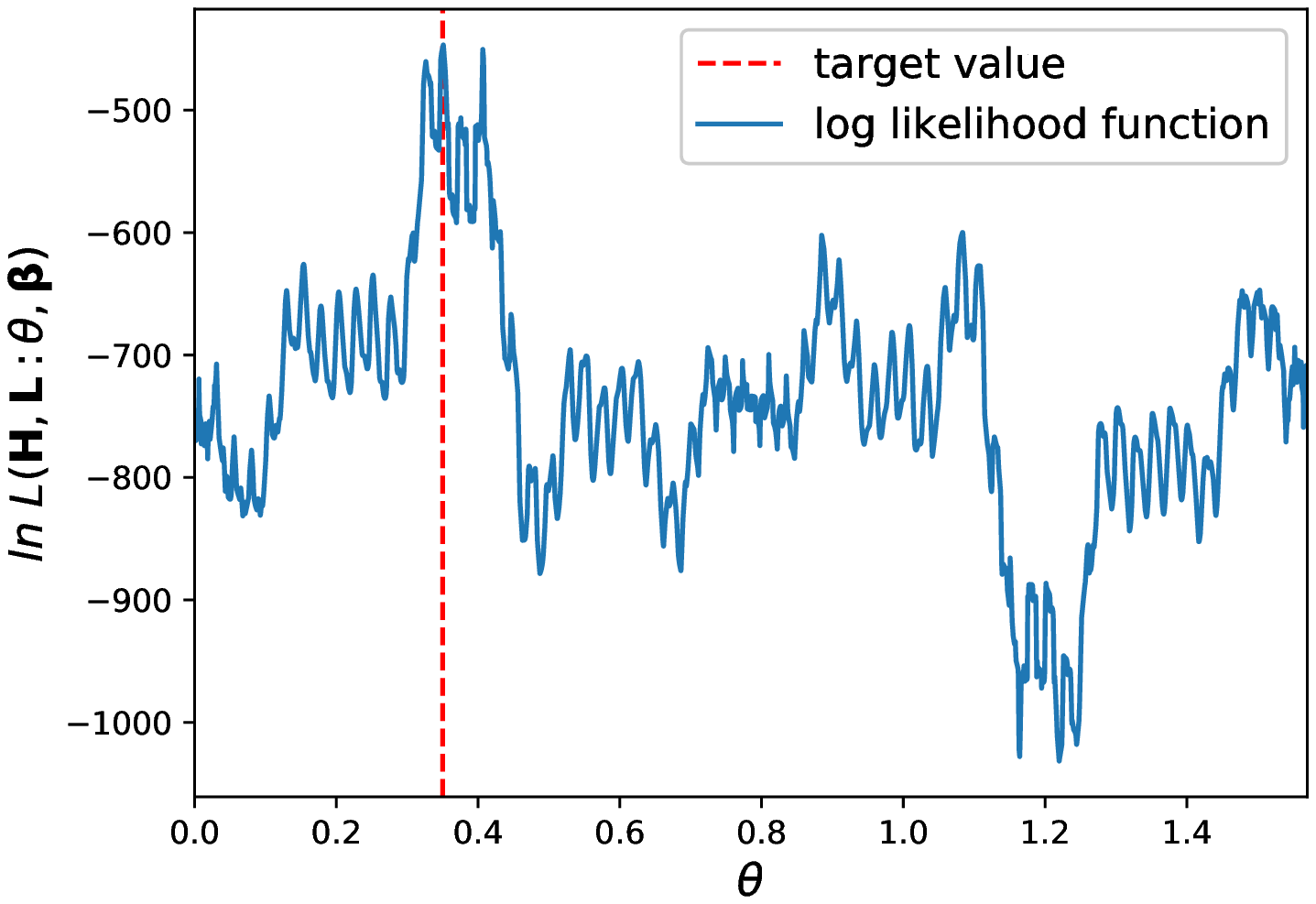} }
		\end{center}
	\end{minipage}
	\caption{(Solid blue) Log likelihood function \eqref{eq:proposedlikelihood}.
		(Dotted red) The target value $\theta_* = 0.35$. }
\end{figure}

Figures~\ref{fig:loglikelihood0} and \ref{fig:loglikelihood1} illustrate the
shape of the log-likelihood function \eqref{eq:proposedlikelihood}, with
Case~${\mathit 1}$ and Case~${\mathit 2}$, respectively;
recall that the other parameters are the same.
They show that, in both cases, the optimal solution of
Eq.~\eqref{likelihood eq Section IV}, or equivalently the maximum point of
the log-likelihood function, almost coincides with the target value
$\theta_* = 0.35$.
That is, certainly the ML estimator is obtained by solving the one-dimensional
maximization problem, without respect to the free parameters $\{c_k\}$.
At the same time, the shape of the likelihood function implies that the
conventional nine-dimensional likelihood function should have a very complicated
landscape and as a result we may easily fail to obtain the ML estimator, which
is indeed the main benefit of the parameter orthogonalization method.

\begin{figure}
	\centering
	\includegraphics[width=\linewidth]{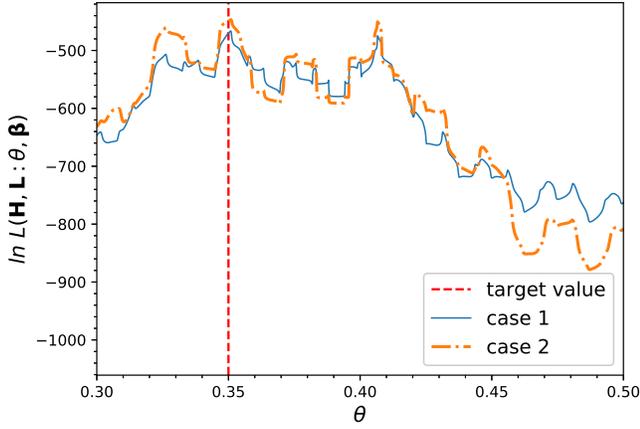}
	\caption{
		Enlarged view of Figs.~\ref{fig:loglikelihood0} and \ref{fig:loglikelihood1}. }
	\label{fig:loglikelihood2}
\end{figure}

In addition to the optimal solution, the other critical points of
$L(\mathbf{H},\mathbf{L} ; \theta, \boldsymbol{\beta})$, i.e., the solutions
to Eq.~\eqref{likelihood eq Section IV}, almost do not depend on $\{c_k\}$.
Figure~\ref{fig:loglikelihood2} shows the enlarged view of
Figs.~\ref{fig:loglikelihood0} and \ref{fig:loglikelihood1} at around the
optimal point.
Obviously, the shape of log-likelihood function changes depending on the
two cases, but notably, their critical points look close to each other.
There are small differences between those points of two functions due to
the relatively small value of $N_{\rm shot}$, but we have observed that they
become small by increasing $N_{\rm shot}$ as predicted by the theory.

\begin{figure}
	\centering
	\includegraphics[width=\linewidth]{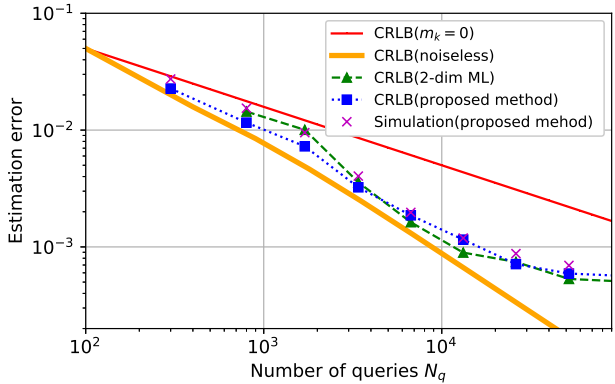}
	\caption{
		Estimation errors of $\theta_*$ vs the total number of queries $N_q$.
		The (red) thin and (yellow) thick lines are the CRLB for the classical
		method ($m_k=0$, i.e., the classical random sampling) and the
		quantum ML method without noise, respectively.
		The (green) dashed and (blue) dotted lines are the CRLB for the
		two-dimension ML estimator assuming the depolarizing noise and the
		proposed ML estimator, respectively.
		The (purple) cross marks are the standard deviation between the
		true value $\theta_*=0.35$ and the estimated values of $\theta$ computed
		using the proposed method. }
	\label{fig:sim}
\end{figure}

Next we study the estimation performance of our ML estimator in the same
setting as above except for the values of $\{c_k\}$. 
Figure~\ref{fig:sim} shows the estimation errors of the target value 
$\theta_*=0.35$,  versus the total number of queries
$N_q=\Sigma_{k=1}^M N_{\rm{shot}}(2m_k + 1)$.
The (red) thin and (yellow) thick lines are the CRLB for the classical method
($m_k=0$, i.e., the classical random sampling) and the quantum ML method
without noise, respectively; the latter decreases the error quadratically faster
than the former, as theoretically proven.
The (green) dashed and (blue) dotted lines are the CRLB for the two-dimensional
ML estimator assuming the depolarizing noise and the proposed ML estimator,
respectively.
Recall that those three ML estimators employ the operating schedule $m_k=2^{k-1}$.
Also, our method makes $N_{\rm{shot}}=50$ measurements for both the Grover
and the ancillary Grover circuits to construct the ML estimator, meaning that the
number of measurements is $100$ for each $m_k$; hence for a fair comparison,
the other ML estimators shown with the yellow solid and green dashed lines are
assumed to make $100$ measurements for each $m_k$. 
In addition, the CRLB of the proposed method is calculated as the 
$(1,1)$ element of the inverse of the Fisher information matrix, 
which does not change before and after the parameter orthogonalization. 
The (purple) cross marks are the standard deviation between the true value $\theta_*=0.35$ and the estimated values of $\theta$ computed using our method which
in this case take $c_k=0.3$ for all $k$.

The first notable point is that the green dashed line (the model assuming 
the depolarization) and the blue dotted line (the model not assuming the 
depolarization) are close with each other.
Considering the fact that the true distribution is now subjected to only 
the depolarizing noise, this result means that our over-parametrized model 
can correctly capture the true distribution.
Note that these CRLBs beat the classical estimation limit (the red thin 
solid line) up to a certain value of $N_q$, as theoretically predicted 
in \cite{tanaka2021amplitude}.
Another important fact is that the estimation errors of the constructed 
estimator (the purple cross marks) well approximate the CRLB. 
That is, the estimator has the asymptotic consistency property, meaning 
that we are successfully solving the optimization problem and accordingly 
obtained the ML estimator almost correctly.
This is clearly thanks to the advantage that the complicated nine-dimensional 
optimization problem now boils down to the one-dimensional one; the complicated 
shape of the likelihood function observed
in Fig.~\ref{fig:loglikelihood2} implies that the ML estimator maximizing 
the nine-dimensional function is hard to obtain, and as a result the gap 
between the purple cross marks and the blue dotted line
can easily become large.


\subsection{Experiment on real quantum device}

Here we show the result of an experiment conducted on IBM Quantum device
``ibm\_kawasaki,'' to study how well our proposed estimator can actually
manage the unidentifiable uncertainty arising in the real device. 
For this purpose we consider the problem of estimating the sum
$S = \sum_{j=0}^{2^n-1} f(j) r(j)$ in the QAE framework 
\cite{suzuki2020,tanaka2021amplitude,QiskitAEwoPE}.
In fact, $S$ can be encoded into the amplitude of a quantum state via the
operator $\mathcal{A}=\mathcal{T}(\mathcal{P} \otimes I_{1})$ as follows;
\begin{equation*}
	\begin{split}
		\mathcal{A}\Ket{0}_n & \Ket{0}
		=\mathcal{T} \sum_j \sqrt{r(j)}\ket{j}_n\ket{0} \\
		&=\sum_j \sqrt{r(j)}\ket{j}_n
		\left(\sqrt{f(j)}\ket{1}+\sqrt{1-f(j)}\ket{0}\right) \\
		&=\sqrt{S} \ket{\tilde{\Psi}_1}\ket{1} + \sqrt{1-S}\ket{\tilde{\Psi}_0}\ket{0},
	\end{split}
\end{equation*}
where $\ket{\tilde{\Psi}_1}=\sum_j \sqrt{r(j) f(j)/S}\ket{j}_n$ and
$\ket{\tilde{\Psi}_0}=\sum_j \sqrt{r(j)(1-f(j))/(1-S)}\ket{j}_n$.
From Eq.~\eqref{eq:defA}, $S=\sin^2\theta$ can be efficiently 
estimated via QAE. 
In this paper, we consider $f(j)=\sin^2(\pi j/10)$ and $r(j)=1/2^n$ for
$\forall j$, with $n=1$; in this case, $\mathcal{P}$ and $\mathcal{T}$
can be implemented using Hadamard and controlled $Y$-rotation gates.
The true value is $\theta = 0.175$ or equivalently $S = 3.03 \times 10^{-2}$.
See \cite{suzuki2020,tanaka2021amplitude,QiskitAEwoPE} for a detailed
description.

\begin{figure}
	\centering
	\includegraphics[width=\linewidth]{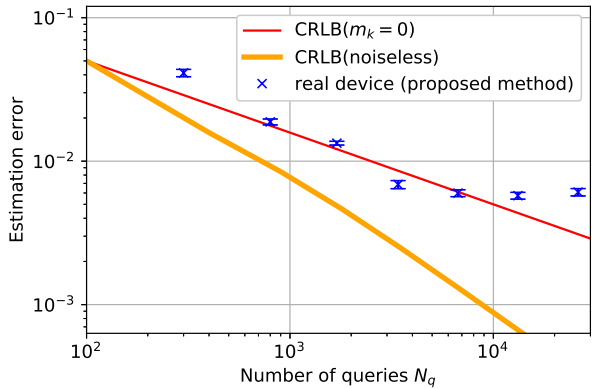}
	\caption{
		Estimation error of $\theta$ vs the total number of queries $N_q$.
		The thin red and thick yellow lines are the CRLB, obtained via the
		classical method, and the quantum ML method without noise, respectively.
		The blue cross marks are the standard deviation between the true value
		$\theta_* = 0.175$ and the estimated values of $\theta$ that is experimentally
		computed using the proposed method.
	}
	\label{fig:real_device}
\end{figure}

For this estimation problem we apply the ML estimator with increasing
number of Grover iterations as $m_k=2^{k-1}$, $(k=1, 2, \ldots, 7)$.
Now our model assumes that a different noise is added to the system for
different value of $m_k$; hence the model contains seven nuisance parameters
$\boldsymbol{\beta}=(\beta_1, \ldots, \beta_7)$, which can be 
removed, however for constructing the ML estimator for $\theta$, using the parameter
orthogonalization method.
Figure~\ref{fig:real_device} shows the result of estimation error versus
the total number of queries, $N_q=\Sigma_{k=1}^M N_{\rm shot}(2m_k + 1)$.
The (red) thin solid line is the CRLB with $m_k=0$ (i.e., classical random
sampling), and the (orange) thick solid line is the CRLB achieved by the
ideal ML method with $m_k=2^k$.
The (blue) cross marks are the standard deviation between the true value
$\theta_* = 0.175$ and the estimated values of $\theta$ obtained via the 
proposed method employing $c_k = 0.3$ for $k=1,\cdots,7$. 
More specifically, the number of measurements is $N_{\rm shot}=50$ for both
$\mathcal{G}^{m_k}$ and $\mathcal{R}\mathcal{G}^{m_k-1}$ to compute one
$\hat{\theta}_{\rm ML}$;
we repeated the same experiment $2,119$ times to compute the standard
deviations (cross marks) and the three-times standard errors (error bars).

In the figure, we roughly see that the cross marks decrease with the
quadratically enhanced scaling $N_q^{-1}$ (nearly parallel to the orange
thick line) up to around $N_q=3\times 10^3$, although there is a
constant-factor overhead. 
This overhead might be due to the additional uncertainty in the device 
as well as the bigger decay rate of the probability amplitude than 
$\kappa = 0.01$ in the previous numerical simulation. 
Nevertheless, the cross mark at around $N_q=3\times 10^3$ is below the 
red line, meaning that the estimator is better than any classical means 
yet only at this point; unfortunately the estimation error does not 
decrease anymore, because excess noise is introduced by further iterating 
the Grover operation. 
In fact, we have confirmed that the noise level of ``ibm\_kawasaki'' is 
comparable to that of ``ibmq\_valenica'' used in the previous study 
\cite{tanaka2021amplitude}, which also exhibited a similar saturation of 
the estimation error. 
These results imply that the one-dimensional maximization problem has been 
solved almost correctly, and the resulting ML estimator works pretty well 
even under the un-identifiable realistic noisy environment. 
Hence we have a perspective that the proposed model, which does not 
incorporate a specific noise characteristic, may be able to capture 
a more complicated larger-dimensional system by increasing the number of 
nuisance parameters $\{\beta_k\}$ and, thanks to the parameter 
orthogonalization, the ML estimator may still be computed almost exactly 
without respect to $\{\beta_k\}$. 
\\


\section{Conclusion}
\label{sec:Conclusion}

Quantum computing can be regarded as a system that encodes and processes
some quantities (parameters) of interest in a real physical device
which are finally retrieved and estimated as precisely as possible.
However, for the time being we will have to play with devices under
unknown noise environment.
The statistical estimation theory provides a useful toolbox for
dealing with such a practical estimation problem, and in our view,
its role will remain or even become bigger when those devices
acquire some level of fault tolerance in the future.
The nuisance parameters method presented in this paper is one such
useful tool.
Recall that what was presented in this paper is not a blind application
of the nuisance parameters method; for instance, we need an additional
quantum circuit (called the ancillary Grover circuit) to apply the
theory and thereby construct an estimator without respect to the
nuisance parameters.
Also, it was somewhat surprising that we can analytically solve the
differential equation \eqref{PDF_org} in our problem; 
as emphasized before, this is indeed a key result obtained in this 
paper, because otherwise (i.e., if the solution has to be numerically 
computed) the parameter orthogonalization procedure may bring a significant 
computational overhead.

Extension to the method formulated within the semi parametric estimation
theory is clearly an important next step of this work.
Also, it should be desirable if such an extension could cover the
problem of quantum phase estimation, which is also an important subroutine
in many quantum algorithms.

\section*{Acknowledgement}
We thank Jun Suzuki for helpful discussions.
This work was supported by the MEXT Quantum Leap Flagship Program through Grants No. JPMXS0118067285 and No. JPMXS0120319794.
The results presented in this paper were obtained in part using an IBM
Quantum quantum computing system as part of the IBM Quantum Network.
The views expressed are those of the authors and do not reflect the
official policy or position of IBM or the IBM Quantum team.

\section*{APPENDIX}

Here we show that, if the new parameters satisfy the differential equation 
(\ref{PDF_org}), then the $(1, k)$-element of the new Fisher information matrix 
$J_\xi$ becomes zero for $k=2,3, \dots, M+1$ \cite{cox1987parameter}.

First, we use the symbol $\tau$ to explicitly represent the variable 
transformation: 
\begin{equation*}
\begin{aligned}
\theta &= \tau_0(\xi_1)=\xi_1, ~ \\
\beta_1 &= \tau_1(\xi_1,\xi_2,\xi_3,\cdots,\xi_{M+1}), \\
\beta_2 &= \tau_2(\xi_1,\xi_2,\xi_3,\cdots,\xi_{M+1}),\\
		& \vdots\\
\beta_M &= \tau_M(\xi_1,\xi_2,\xi_3,\cdots,\xi_{M+1}).
\end{aligned}
\end{equation*}
Also, to make the notation simple, we define 
$f = \log p(x ;\theta, \boldsymbol{\beta})$ and 
$g = \log p(x ;\boldsymbol{\xi})$. 
We now calculate $(J_\xi)_{1,k}$ for $k=2,3, \dots, M+1$ using 
the chain rule as follows:
\begin{widetext}
$$
\begin{aligned}
(J_\xi)_{1,k} &= {\mathbb E}\Big[ \frac{\partial g}
		{\partial \xi_1}
		\frac{\partial g}
		{\partial \xi_k}
		\Big]   
= {\mathbb E}\Big[ 
\Big(
\frac{\partial f } {\partial \theta} \frac{\partial\tau_0 } {\partial \xi_1} +
\sum_{h=2}^{M+1} \frac{\partial f } {\partial \beta_{h-1}} \frac{\partial\tau_{h-1} } {\partial \xi_1}
\Big) \Big(
\frac{\partial f } {\partial \theta} \frac{\partial\tau_0 } {\partial \xi_k} +
\sum_{i=2}^{M+1} \frac{\partial f } {\partial \beta_{i-1}} \frac{\partial\tau_{i-1} } {\partial \xi_k}
\Big) 	\Big] \\   
&= {\mathbb E}\Big[ 
\Big(
\frac{\partial f } {\partial \theta}  +
\sum_{h=2}^{M+1} \frac{\partial f } {\partial \beta_{h-1}} \frac{\partial\tau_{h-1} } {\partial \xi_1}
\Big) \Big(
\sum_{i=2}^{M+1} \frac{\partial f } {\partial \beta_{i-1}} \frac{\partial\tau_{i-1} } {\partial \xi_k}
\Big) 	\Big]   
= \sum_{i=2}^{M+1} \frac{\partial\tau_{i-1} } {\partial \xi_k} {\mathbb E}\Big[ 
\Big(
\frac{\partial f } {\partial \theta}  +
\sum_{h=2}^{M+1} \frac{\partial f } {\partial \beta_{h-1}} \frac{\partial\tau_{h-1} } {\partial \xi_1}
\Big) 
\frac{\partial f } {\partial \beta_{i-1}} 
\Big] \\  
&= \sum_{i=2}^{M+1} \frac{\partial\tau_{i-1} } {\partial \xi_k} 
\Big(
{\mathbb E}\Big[ 
\frac{\partial f } {\partial \theta} \frac{\partial f } {\partial \beta_{i-1}} \Big]
+
\sum_{h=2}^{M+1} {\mathbb E}\Big[ 
\frac{\partial f } {\partial \beta_{i-1}} 
\frac{\partial f }{\partial \beta_{h-1}} 
\Big] \frac{\partial\tau_{h-1} } {\partial \xi_1}
\Big) 
= \sum_{i=2}^{M+1} \frac{\partial\tau_{i-1} }  {\partial \xi_k} \Big(
J_{1,i} +
\sum_{h=2}^{M+1} J_{i,h} \frac{\partial\tau_{h-1} } {\partial \xi_1}
\Big).  
\end{aligned}
$$
\end{widetext}
Therefore if 
$$
J_{1,i} +
\sum_{h=2}^{M+1} J_{i,h} \frac{\partial\tau_{h-1} } {\partial \xi_1}
=0
$$
for $i=2,3, \dots, M+1$, then we have $(J_\xi)_{1,k} =0$.

\bibliography{Manuscript}

\end{document}